\documentclass[aps,prl,twocolumn,showpacs,floatfix,superscriptaddress,letterpaper,longbibliography]{revtex4-2}
\usepackage{graphicx}
\usepackage{bm}
\usepackage{amsmath,amsfonts}
\usepackage{amsbsy}
\usepackage[utf8]{inputenc}
\DeclareUnicodeCharacter{03BC}{\mbox{$\mu$}}
\DeclareUnicodeCharacter{2009}{ }

\begin{document}

\title{Two-fluid dynamics of one-dimensional quantum liquids in the
  absence of Galilean invariance}

\author{K. A. Matveev}

\affiliation{Materials Science Division, Argonne National Laboratory,
  Argonne, Illinois 60439, USA}

\author{A. V. Andreev}

\affiliation{Department of Physics, University of Washington, Seattle,
  Washington 98195, USA}

\date{May 8, 2019}

\begin{abstract}

  Luttinger liquid theory of one-dimensional quantum systems ignores
  exponentially weak backscattering of particles. This endows
  Luttinger liquids with superfluid properties.  The corresponding
  two-fluid hydrodynamic description available at present applies only
  to Galilean-invariant systems, whereas most experimental
  realizations of one-dimensional quantum liquids lack Galilean
  invariance.  Here we develop the two-fluid theory of such quantum
  liquids.  In the low-frequency limit the theory reduces to
  single-fluid hydrodynamics.  However, the absence of Galilean
  invariance brings about three new transport coefficients.  We obtain
  expressions for these coefficients in terms of the backscattering
  rate.
  
\end{abstract}
\maketitle

\emph{Introduction.}---\,One-dimensional quantum liquids are commonly
described in the framework of the so-called Luttinger liquid theory
\cite{haldane_luttinger_1981, giamarchi_quantum_2004}.  In the
simplest form of this theory the elementary excitations are
non-interacting bosons.  In this approximation the system does not
equilibrate.  Interactions between the excitations arise from
corrections to the Luttinger liquid Hamiltonian, which are irrelevant
perturbations in the renormalization group sense.  The resulting
scattering processes give rise to relaxation of the system to thermal
equilibrium.  The irrelevant perturbations scale as powers of energy,
and thus the corresponding relaxation rate $\tau_{\rm ex}^{-1}$ scales
as a power of temperature $T$.  Importantly, the collisions generated
by these irrelevant perturbations conserve not only the number of
particles, momentum, and energy of the system, but also an additional
quantity $J$, which was first introduced in
Ref.~\cite{haldane_luttinger_1981}.  It was shown recently
\cite{matveev_second_2017, matveev_hybrid_2018,
  matveev_propagation_2018} that as a result Luttinger liquids behave
as superfluids and, similarly to superfluid $^4$He
\cite{landau_theory_1941, khalatnikov_introduction_2000}, can be
described by two-fluid hydrodynamics.

The reason the system behaves as a superfluid can be understood as
follows.  The physical meaning of the conserved quantity $J$ is most
transparent for quantum liquids composed of fermions, in which case
$J$ is the difference of the numbers of right- and left-moving
particles.  At $J\neq0$ even in the absence of elementary excitations
the quantum liquid has a finite momentum $p_F J$ and velocity
$u_0=p_F J/mN$.  Here $p_F$ is the Fermi momentum, and $m$ and $N$ are
the mass and number of particles.  At non-zero temperature the
elementary excitations form a gas that moves with its own velocity
$u_{\rm ex}$.  Thus, similar to superfluid $^4$He, Luttinger liquid
exhibits two types of macroscopic motion, and its flow should be
described by two-fluid hydrodynamics.

Luttinger liquid theory accounts only for the low-energy excitations
of the system.  On the other hand, the scattering processes involving
excitations with energies of order of the bandwidth $D$ can
backscatter fermions and thus violate conservation of $J$.  The most
efficient processes of this type involve holes near the bottom of the
band \cite{lunde_three-particle_2007}, and thus have a rate that is
exponentially small at low temperatures, $\tau^{-1}\propto e^{-D/T}$
\cite{micklitz_transport_2010, matveev_equilibration_2012,
  matveev_scattering_2012}.  Because of that the superfluid behavior
of one-dimensional quantum liquids is limited to the frequency range
$ \tau^{-1} \ll \omega \ll \tau_{\rm ex}^{-1}$.  At the lowest
frequencies $\omega\ll\tau^{-1}$, backscattering processes lead to
equilibration of the velocities, $u_{\rm ex}-u_0\to0$, and the quantum
liquid behaves as a normal fluid.

Dynamics of one-dimensional quantum liquids at frequencies
$\omega\ll \tau_{\rm ex}^{-1}$ is described by the two-fluid
hydrodynamic theory \cite{matveev_second_2017, matveev_hybrid_2018,
  matveev_propagation_2018}, which was obtained by adapting the
Landau's theory of superfluidity of liquid $^4$He
\cite{landau_theory_1941, khalatnikov_introduction_2000} to one
dimension.  An important limitation of this approach is the assumption
that the system is Galilean-invariant.  On the other hand, for quantum
liquids realized in solid state systems the underlying crystalline
lattice plays an important role and usually leads to violation of
Galilean invariance.  Development of two-fluid hydrodynamic theory of
one-dimensional quantum liquids in this regime is our main goal.

An important property of Galilean-invariant systems is that the
particle number current $j_n$ is given by the momentum density $p$
divided by the particle mass $m$.  Therefore, the collisions between
the particles, which conserve momentum, do not affect $j_n$.  This is
in contrast with the energy current $j_\varepsilon$, which in addition
to a contribution proportional to $p$ contains a dissipative
contribution, typically expressed in terms of the thermal conductivity
and temperature gradient.  In the absence of Galilean invariance $j_n$
is not uniquely defined by momentum density and, similar to
$j_\varepsilon$, has a dissipative component.  The particle and energy
currents appear in response to gradients of chemical potential $\mu$
and temperature $T$.  In a linear approximation the dissipative parts
of $j_n$ and $j_\varepsilon$ are related to the gradients
$\partial_x\mu$ and $\partial_x T$ by a matrix of four transport
coefficients, similar to thermoelectric coefficients in conductors
\cite{abrikosov_fundamentals_1988}.  Our two-fluid theory enables us
to express these transport coefficients in terms of the relaxation
time $\tau$.

\emph{Luttinger liquid.}---For simplicity, we consider a quantum
liquid of spinless particles.  In the Luttinger liquid approximation
it is described in terms of the bosonic fields $\varphi(x)$ and
$\theta(x)$, satisfying the commutation relation
\begin{equation}
  \label{eq:commutation_relations}
  [\varphi(x),\partial_{x'}\theta(x')]=i\pi\delta(x-x').
\end{equation}
The fields are subject to the  boundary conditions
\begin{equation}
  \label{eq:boundary_conditions}
  \varphi(L)=\varphi(0),
  \quad
  \theta(L)=\theta(0)-\pi J,
\end{equation}
where $L$ is the size of the system.  The second condition formally
defines the conserved quantity $J$, which for systems of fermions is
the difference of the numbers of right- and left-moving particles.  It
will be convenient to also define a field $\vartheta(x)$, which
satisfies periodic boundary conditions $\vartheta(L) = \vartheta(0)$
and is related to $\theta(x)$ by
\begin{equation}
  \label{eq:kappa_definition}
  \partial_x\theta=-\frac{\chi}{\hbar}+\partial_x\vartheta,
  \quad
  \chi=\pi\hbar\frac{J}{L}.
\end{equation}
Here $\hbar$ is the Planck's constant, while $\chi$ has the dimension of
momentum and is proportional to the density of the conserved quantity
$J$.

The momentum of the Luttinger liquid can be expressed in terms of the
bosonic fields as \cite{haldane_luttinger_1981,onefootnote}
\begin{equation}
  \label{eq:Momentum_operator}
  P=-\hbar\int dx \bigg(n+\frac{\partial_x\varphi}{\pi} \bigg)
  \partial_x\theta
  =N\chi -\frac{\hbar}{\pi}\int dx\, \partial_x\varphi\, \partial_x\vartheta.
\end{equation}
Here we denote the total number of particles by $N$, while $n=N/L$ is
the average density.  The first form of
Eq.~(\ref{eq:Momentum_operator}) is clear from the physical meaning of
the bosonic fields: $\partial_x\varphi(x)/\pi$ is the density of
particles at point $x$, measured from $n$, whereas
$-\hbar\,\partial_x\theta(x)$ is the momentum of the liquid per
particle.  In the second form of Eq.~(\ref{eq:Momentum_operator}) for
a liquid of spinless fermions $N\chi$ can be expressed in terms of
the Fermi momentum $p_F=\pi\hbar n$ as $p_FJ$ and represents the
momentum of the ground state with unequal numbers of the right- and
left-moving particles.  The remaining integral of
$-(\hbar/\pi) \partial_x\varphi\, \partial_x\vartheta$ accounts for
the momentum of the elementary excitations.

The Hamiltonian of the Luttinger liquid can be presented in the form
\cite{haldane_luttinger_1981, giamarchi_quantum_2004}
\begin{eqnarray}
  \label{eq:Hamiltonian}
  H&=&\int dx
    \bigg\{
      \frac{\hbar v}{2\pi}\left[K(\partial_x\theta)^2
         +\frac{1}{K}(\partial_x\varphi)^2\right]
\nonumber\\
  &&\qquad\quad+\alpha_\theta^{} \partial_x\varphi(\partial_x\theta)^2
  +\alpha_\varphi (\partial_x\varphi)^3+\ldots
    \bigg\},
\end{eqnarray}
where $v$ is the velocity of the elementary excitations and $K$ is the
Luttinger liquid constant.  The coefficients $\alpha_\theta$ and
$\alpha_\varphi$ can be expressed in terms of the dependences of $v$
and $K$ on density as \cite{matveev_scattering_2012}
\begin{equation}
\label{eq:alpha_relations}
  \alpha_\theta^{} = \frac{\hbar}{ 2\pi^2}\,  \partial_n\left(v K\right),
  \quad
  \alpha_\varphi = \frac{\hbar}{6\pi^2}\, \partial_n \left( \frac{v}{K}\right).
\end{equation}
We omitted from the Hamiltonian (\ref{eq:Hamiltonian}) terms with
scaling dimensions exceeding three.

In most applications the Hamiltonian of the Luttinger liquid is
approximated by the first line of Eq.~(\ref{eq:Hamiltonian}).  In this
case the elementary excitations are noninteracting bosons.  The energy
of the boson with momentum $q$ is $\epsilon_q=v|q|$
\cite{haldane_luttinger_1981, giamarchi_quantum_2004}.  We will need
to account for a correction to this expression arising at small but
finite $\chi$.  To this end we substitute
Eq.~(\ref{eq:kappa_definition}) into the cubic term in
Eq.~(\ref{eq:Hamiltonian}) proportional to $\alpha_\theta$.  To linear
order in $\chi$ this generates a correction to the quadratic
Hamiltonian in the first line of Eq.~(\ref{eq:Hamiltonian})
proportional to the momentum of the elementary excitations.  As a
result, the energy spectrum takes the form
\begin{equation}
  \label{eq:quasiparticle_energy}
  \epsilon_q=v|q|+\frac{\chi}{m^*}q.
\end{equation}
Here we have introduced the effective mass
\begin{equation}
  \label{eq:m*}
  m^*=\frac{\pi\hbar}{\partial_n(vK)}.
\end{equation}
In the Galilean-invariant case the parameters $v$ and $K$ are not
independent.  Specifically, $vK=v_F$ where the Fermi velocity
$v_F=\pi\hbar n/m$.  In this case the effective mass (\ref{eq:m*})
coincides with the mass $m$ of the particles comprising the
Luttinger liquid.  Equation (\ref{eq:quasiparticle_energy}) can then
be interpreted as the Galilean transformation of the original spectrum
$\epsilon_q=v|q|$ to the frame moving with velocity $\chi/m$.

\emph{Thermal equilibrium.}---We now consider an equilibrium state of
the Luttinger liquid.  It is described by the number of particles $N$,
the value of $J$, and the boson occupation numbers $N_q$, which take
the usual Bose form
\begin{equation}
  \label{eq:Bose_distribution}
  N_q=\left[\exp\left(\frac{\epsilon_q-uq}{T}\right)-1\right]^{-1}.
\end{equation}
The parameter $u$ appears as a consequence of the conservation of
momentum and can be thought of as the velocity of the gas of
excitations.  It is worth noting that the occupation numbers
\eqref{eq:Bose_distribution} implicitly depend on $N$ and $J$, because
the velocity of the excitations $v$ in
Eq.~\eqref{eq:quasiparticle_energy} is a function of density, while
$\chi\propto J$.

Using the distribution function \eqref{eq:Bose_distribution} it is
straightforward to obtain the thermodynamic properties of the system
to leading order in $T/D$.  Below we will need expressions for the
energy and momentum densities of the Luttinger liquid.  We will limit
our consideration to low-velocity flows and neglect contributions
beyond linear order in $u$ and $\chi$.  The energy density
$\varepsilon$ is obtained by adding contributions $\epsilon_qN_q$ of
all excitations, resulting in
\begin{equation}
  \label{eq:energy_density}
  \varepsilon=\frac{\pi T^2}{6\hbar v}.
\end{equation}
We obtain the momentum density $p$ using the second form of
Eq.~\eqref{eq:Momentum_operator}, in which the second term is given by
the sum of the contributions $qN_q$ of all bosonic states.  This
yields
\begin{equation}
  \label{eq:Momentum_density}
  p=n\chi+\rho_{\rm ex}\left(u-\frac{\chi}{m^*}\right),
  \quad
  \rho_{\rm ex}=\frac{\pi T^2}{3\hbar v^3}.
\end{equation}
We earlier interpreted $u$ as the velocity of the gas of bosonic
excitations of the Luttinger liquid.  Given that the corresponding
contribution to the momentum density is 
$\rho_{\rm ex}u$, one can interpret $\rho_{\rm
ex}$ as the mass density of the gas of excitations. 

\emph{Hydrodynamics of Luttinger liquids.}---Let us now consider the
dynamics of the Luttinger liquid at time scales much longer than
$\tau_{\rm ex}$.  In this regime the liquid is in local thermal
equilibrium at every point.  As a result, its state is fully described
by the densities of the four conserved quantities: number of
particles, energy, momentum, and $J$.  Time evolution of these
densities is described by four continuity equations expressing the
conservation laws:
\begin{eqnarray}
  \label{eq:continuity_n}
  \partial_t n + \partial_xj_n &=&0,
  \\
  \label{eq:continuity_epsilon}
  \partial_t \varepsilon + \partial_xj_\varepsilon &=&0,
  \\
  \label{eq:continuity_p}
  \partial_t p + \partial_xj_p &=&0,
  \\
  \label{eq:continuity_kappa}
  \partial_t \chi + \partial_xj_\chi &=&0,
\end{eqnarray}
where we used $\chi$ defined in \eqref{eq:kappa_definition} instead
of the density $J/L$.  In order to complete the hydrodynamic
description of the system we need to evaluate the currents $j_n$,
$j_\varepsilon$, $j_p$, and $j_\chi$.  We will restrict ourselves to
terms up to linear order in small parameters $\chi(x)$, $u(x)$, and
$\delta n(x)=n(x)-N/L$.

To evaluate the particle current $j_n$ we write the Heisenberg
equation of motion for the operator of particle density
$n(x)=N/L+\partial_x\varphi(x)/\pi$.  This yields the evolution
equation \eqref{eq:continuity_n} in the operator form with
\begin{equation}
  \label{eq:j_n_operator}
  j_n(x)=\frac{i}{\pi\hbar}[\varphi(x),H]
  =-\frac{vK}{\pi}\partial_x\theta
  -\frac{2\alpha_\theta}{\hbar}\partial_x\varphi\,\partial_x\theta.
\end{equation}
Next we substitute $\partial_x\theta$ in the form
\eqref{eq:kappa_definition} and evaluate the expectation value of the
operator \eqref{eq:j_n_operator} in the equilibrium state of the
Luttinger liquid.  Introducing a new effective mass 
\begin{equation}
  \label{eq:m0}
  m_0=\frac{\pi\hbar n}{vK}
\end{equation}
and noting that the operator
$(-\hbar/\pi)\partial_x\varphi\:\!\partial_x\vartheta$ is the momentum
density of the bosonic excitations (see above), we obtain
\begin{equation}
  \label{eq:j_n}
  j_n=\frac{n\chi}{m_0}
  +\frac{\rho_{\rm ex}}{m^*}\left(u-\frac{\chi}{m^*}\right).
\end{equation}
For Galilean-invariant systems $vK=v_F$, and thus $m_0=m$.  In this
case $j_n$ must coincide with the ratio of momentum density $p$ and
the particle mass $m$.  This is easily verified by substituting
$m_0=m^*=m$ into Eqs.~\eqref{eq:Momentum_density} and \eqref{eq:j_n}.

The above procedure can be extended to the evaluation of the remaining
three currents.  To obtain $j_\varepsilon$ one can write the equation
of motion for the operator of Hamiltonian density, keeping both the
quadratic and cubic terms in Eq.~\eqref{eq:Hamiltonian}.  After
thermal averaging one obtains
\begin{equation}
  \label{eq:j_varepsilon}
  j_\varepsilon=2\varepsilon u
             +\varepsilon\,\frac{n\partial_nv}{v}\,\frac{\chi}{m_0}.
\end{equation}
The evaluation of the momentum current $j_p$ starts with the equation
of motion for the operator of momentum density
$-\hbar(n+\partial_x\varphi/\pi)\partial_x\theta$,
cf.\ Eq.~\eqref{eq:Momentum_operator}.  The result is
\begin{equation}
  \label{eq:j_p}
  j_p=m_0v^2\delta n+\varepsilon\,\frac{\partial_n(nv)}{v}.
\end{equation}
Finally, the current $j_\chi$ is obtained by averaging the equation
of motion for the operator $\partial_x\theta$ and using
Eq.~\eqref{eq:kappa_definition}.  This yields
\begin{equation}
  \label{eq:j_kappa}
  j_\chi=\frac{m_0v^2}{n}\delta n +\varepsilon\,\frac{\partial_nv}{v}.
\end{equation}
It is easy to show that the above expressions for $j_p$ and $j_\chi$
coincide with the pressure $\Pi$ and chemical potential $\mu$ of the
Luttinger liquid, respectively.

Compared to conventional hydrodynamics, the above theory includes an
additional equation \eqref{eq:continuity_kappa} originating from the
conservation of $J$.  In the Galilean-invariant case the resulting
theory was interpreted \cite{matveev_second_2017, matveev_hybrid_2018,
  matveev_propagation_2018} as two-fluid hydrodynamics, fully
analogous to that of superfluid $^4$He in three dimensions
\cite{landau_theory_1941, khalatnikov_introduction_2000}.  The normal
component of the fluid in this analogy is the gas of excitations
characterized by velocity $u$, while the superfluid component
describes the Fermi surface and moves with velocity $\chi/m$.  In
the absence of Galilean invariance the full set of two-fluid
hydrodynamic equations of the Luttinger liquid is given by the four
continuity equations
\eqref{eq:continuity_n}--\eqref{eq:continuity_kappa} along with the
constitutive relations \eqref{eq:j_n}--\eqref{eq:j_kappa}.  The latter
were derived to linear order in the deviations from the static
equilibrium state, in which $\delta n$, $\chi$ and $u$ vanish.  To
this approximation the system is described by two effective masses,
$m^*$ and $m_0$.  The Galilean invariant limit is recovered at
$m^*=m_0=m$.

\emph{Hydrodynamics of one-dimensional quantum liquids.}---As
discussed above, Luttinger liquid approximation neglects the
exponentially weak backscattering processes, which relax $J$ to its
equilibrium value.  The latter is determined by the velocity $u$ and
can be found as follows.  To leading order at $T\to0$ the value of $J$
is determined by the Gibbs distribution $w_J\propto e^{-(E-uP)/T}$,
where $E=\pi\hbar v KJ^2/2L$ is the ground state energy of the
Luttinger liquid, obtained by substituting
Eq.~\eqref{eq:kappa_definition} into the first line of
Eq.~\eqref{eq:Hamiltonian}, and the momentum $P=\pi\hbar N J/L$ is
obtained from Eq.~\eqref{eq:Momentum_density}.  The maximum of this
distribution gives the equilibrium value $J=uN/vK$ or, equivalently,
$\chi=m_0u$.

At time scales much longer than $\tau_{\rm ex}$ the deviation of the
system from equilibrium is described by $\chi- m_0u\neq0$.  For
small deviations, relaxation to equilibrium proceeds with the
exponentially small rate $\tau^{-1}$ \cite{micklitz_transport_2010,
  matveev_equilibration_2012, matveev_scattering_2012}, following the
usual relaxation law
\begin{equation}
  \label{eq:relaxation}
  \frac{d}{dt}(\chi-m_0u) = -\frac{1}{\tau}(\chi-m_0u).
\end{equation}
The relaxation processes obey the remaining three conservation laws.
In particular, the momentum density is conserved, $dp/dt=0$.  Using
the expression \eqref{eq:Momentum_density}, we obtain a linear
relation between the $d\chi/dt$ and $du/dt$.  This enables us to
find $d\chi/dt$.  We then find the time evolution equation for
$\chi$ by substituting $d\chi/dt$ into the right-hand side of
Eq.~\eqref{eq:continuity_kappa},
\begin{equation}
  \label{eq:kappa_evolution}
  \partial_t \chi +\partial_x
  j_\chi=-\frac{1}{\tau}\,\frac{\rho_{\rm ex}}{m_0n}(\chi-m_0u).
\end{equation}
Here we accounted only for the leading at small $T$ contribution in
the right-hand side.

Equation \eqref{eq:kappa_evolution} completes our generalizaiton of
two-fluid hydrodynamics of one-dimensional quantum liquids to the
non-Galilean-invariant case.  The dynamics of the system at
frequencies $\omega\ll\tau_{\rm ex}^{-1}$ is fully described by Eqs.~\eqref{eq:continuity_n}--\eqref{eq:continuity_p} and
\eqref{eq:kappa_evolution} along with the constitutive relations
\eqref{eq:j_n}--\eqref{eq:j_kappa}.  At
$\tau^{-1}\ll\omega\ll\tau_{\rm ex}^{-1}$ the right-hand side of
Eq.~\eqref{eq:kappa_evolution} can be neglected and the system is
described by
Eqs.~\eqref{eq:continuity_n}--\eqref{eq:continuity_kappa}.  In this
regime the system supports two sound modes, similar to the
Galilean-invariant case \cite{matveev_second_2017,
  matveev_hybrid_2018, matveev_propagation_2018}.  On the other hand,
at low frequencies $\omega\ll \tau^{-1}$ qualitatively new physics
arises.

\emph{Dissipation in non-Galilean-invariant fluids.}---In the
low-frequency limit the backscattering processes are very effective at
bringing the system to equilibrium characterized by only three
conserved quantities: number of particles, energy, and momentum. The
quantity $J$ is no longer conserved; in a uniform fluid it takes the
equilibrium value $uN/vK$.  In a non-uniform fluid, the system
approaches a local equilibrium at every point in space, with
$\chi(x)=m_0u(x)$.  However, full local equilibrium cannot be
achieved in a non-uniform system, i.e., the deviation from equilibrium
$\delta \chi=\chi-m_0u$ acquires a finite value proportional to
the gradients of physical parameters of the system, such as the
temperature and chemical potential.  As a result, the system behaves
as a conventional fluid described the hydrodynamic equations
\eqref{eq:continuity_n}--\eqref{eq:continuity_p}, with the
constitutive relations containing dissipative contributions
proportional to $\partial_xT$ and $\partial_x\mu$.

We now apply our two-fluid description to evaluate dissipative
components of the currents.  We start by using
Eq.~\eqref{eq:Momentum_density} to bring $\delta\chi$ to the form
$\delta\chi = (m_0/\rho_{\rm ex}) \{[n+\rho_{\rm ex}
(m_0^{-1}-{m^*}^{-1})] \chi-p\}$.  We then use
Eqs.~\eqref{eq:continuity_p} and \eqref{eq:kappa_evolution} to express
$\partial_t\delta\chi$ as a linear combination of $\delta\chi$,
$\partial_x j_p=\partial_x\Pi$, and $\partial_xj_\chi=\partial_x\mu$.
At low frequencies the time derivative
$\partial_t\delta\chi \sim\omega\delta\chi \ll\delta\chi/\tau$ and can
be neglected.  This enables us to express $\delta\chi$ as
\begin{equation}
  \label{eq:delta_kappa}
  \delta\chi = \tau\frac{m_0s}{\rho_{\rm ex}}\partial_xT
                -\tau\left(1-\frac{m_0}{m^*}\right)\partial_x\mu.
\end{equation}
Here we kept only the leading at $T\to0$ terms and used the
thermodynamic relation $\partial_x\Pi=n\partial_x\mu+s\partial_xT$ to
express the gradient of pressure in terms of the gradients of
temperature and chemical potential, with $s=\pi T/3\hbar v$ being the
entropy density of the Luttinger liquid.

The particle and energy currents given by Eqs.~\eqref{eq:j_n} and
\eqref{eq:j_varepsilon} are linear in $u$ and $\chi$.  One can get
further insight into the physics associated with these currents by
expressing them instead in terms of the momentum density
\eqref{eq:Momentum_density} and $\delta\chi=\chi-m_0u$.  Due to
conservation of momentum, the parts of $j_n$ and $j_\varepsilon$
proportional to $p$ are not affected by the relaxation processes and
represent the equilibrium contributions to the currents.  The
components of the currents proportional to $\delta\chi$ correspond to
dissipative contributions.  In the low-frequency limit, the latter can
be expressed in terms of the gradients of temperature and chemical
potential with the aid of Eq.~\eqref{eq:delta_kappa}.  This yields
\begin{eqnarray}
  \label{eq:j_n_dissipative}
  j_n&=&\frac{p}{m_0}
     -\gamma_{11}\partial_x\mu
     -\gamma_{12}\frac{\partial_xT}{T},
\\
  \label{eq:j_varepsilon_dissipative}
  j_\varepsilon&=&\frac{p}{m_0}\frac{\partial_n(n^2v)}{n^2v}\varepsilon
  -\gamma_{21}\partial_x\mu
  -\gamma_{22}\frac{\partial_xT}{T},
\end{eqnarray}
where
\begin{eqnarray*}
  \label{eq:gammas}
  &&\gamma_{11}=\tau\rho_{\rm
     ex}\bigg(\frac{1}{m^*}-\frac{1}{m_0}\bigg)^2,
     \quad
     \gamma_{22}=2\tau\varepsilon v^2,
  \\
  &&\gamma_{12}=\gamma_{21}
  =2\tau\varepsilon \bigg(\frac{1}{m^*}-\frac{1}{m_0}\bigg).  
\end{eqnarray*}
Hydrodynamics of one-dimensional quantum liquids at
$\omega\ll\tau^{-1}$ is fully described by the set of three continuity
equations \eqref{eq:continuity_n}--\eqref{eq:continuity_p} along with
the expressions \eqref{eq:j_p}, \eqref{eq:j_n_dissipative}, and
\eqref{eq:j_varepsilon_dissipative} for the currents.

In the Galilean-invariant case, $m^*=m_0=m$, the dissipative
contribution to the particle current vanishes, and
Eq.~\eqref{eq:j_n_dissipative} recovers the expected relation
$j_n=p/m$.  The energy current contains both the equilibrium and
dissipative contributions, with the latter defining the thermal
conductivity of the liquid $\kappa=\gamma_{22}/T$, cf.\
Ref.~\cite{degottardi_electrical_2015}.  In the absence of Galilean
invariance the dissipation is described by the matrix of four
coefficients $\gamma_{ij}$.  The latter relates dissipative components
of $j_n$ and $j_\varepsilon$ to $\partial_x\mu$ and $\partial_xT/T$,
and is analogous to the martix of thermoelectric coefficients in
conductors \cite{abrikosov_fundamentals_1988}.  The coefficient
$\gamma_{11}$ is analogous to electrical conductivity.  The
off-diagonal martix elements satisfy the Onsager relation,
$\gamma_{12}=\gamma_{21}$.

\emph{Summary.}---We have developed a hydrodynamic theory of
non-Galilean-invariant one-dimensional quantum liquids, which applies
at frequencies $\omega\ll\tau_{\rm ex}^{-1}$.  The fluid is described
by the three usual continuity equations
\eqref{eq:continuity_n}--\eqref{eq:continuity_p}, expressing
conservation of particle number, energy, and momentum, and the
additional evolution equation \eqref{eq:kappa_evolution}.  At small
deviations from equilibrium the dynamics of the system is
characterized by two effective masses, $m^*$ and $m_0$, both of which
coincide with the particle mass in the Galilean-invariant case.  At
the lowest frequencies, $\omega\ll\tau^{-1}$, our theory reduces to
ordinary hydrodynamics.  However, the absence of Galilean invariance
results in the emergence of new transport coefficients, analogous to
conductivity and thermoelectric coefficients.

\begin{acknowledgments}
  
  The authors are grateful to I.~L. Aleiner and W. DeGottardi for
  insightful discussions.  Work at Argonne National Laboratory was
  supported by the U.S. Department of Energy, Office of Science,
  Materials Sciences and Engineering Division.  Work at the University
  of Washington was supported by the U.S.  Department of Energy Office
  of Science, Basic Energy Sciences under Award No. DE-FG02-07ER46452.

\end{acknowledgments}


\begin{thebibliography}{14}%
\makeatletter
\providecommand \@ifxundefined [1]{%
 \@ifx{#1\undefined}
}%
\providecommand \@ifnum [1]{%
 \ifnum #1\expandafter \@firstoftwo
 \else \expandafter \@secondoftwo
 \fi
}%
\providecommand \@ifx [1]{%
 \ifx #1\expandafter \@firstoftwo
 \else \expandafter \@secondoftwo
 \fi
}%
\providecommand \natexlab [1]{#1}%
\providecommand \enquote  [1]{``#1''}%
\providecommand \bibnamefont  [1]{#1}%
\providecommand \bibfnamefont [1]{#1}%
\providecommand \citenamefont [1]{#1}%
\providecommand \href@noop [0]{\@secondoftwo}%
\providecommand \href [0]{\begingroup \@sanitize@url \@href}%
\providecommand \@href[1]{\@@startlink{#1}\@@href}%
\providecommand \@@href[1]{\endgroup#1\@@endlink}%
\providecommand \@sanitize@url [0]{\catcode `\\12\catcode `\$12\catcode
  `\&12\catcode `\#12\catcode `\^12\catcode `\_12\catcode `\%12\relax}%
\providecommand \@@startlink[1]{}%
\providecommand \@@endlink[0]{}%
\providecommand \url  [0]{\begingroup\@sanitize@url \@url }%
\providecommand \@url [1]{\endgroup\@href {#1}{\urlprefix }}%
\providecommand \urlprefix  [0]{URL }%
\providecommand \Eprint [0]{\href }%
\providecommand \doibase [0]{https://doi.org/}%
\providecommand \selectlanguage [0]{\@gobble}%
\providecommand \bibinfo  [0]{\@secondoftwo}%
\providecommand \bibfield  [0]{\@secondoftwo}%
\providecommand \translation [1]{[#1]}%
\providecommand \BibitemOpen [0]{}%
\providecommand \bibitemStop [0]{}%
\providecommand \bibitemNoStop [0]{.\EOS\space}%
\providecommand \EOS [0]{\spacefactor3000\relax}%
\providecommand \BibitemShut  [1]{\csname bibitem#1\endcsname}%
\let\auto@bib@innerbib\@empty
%</preamble>
\bibitem [{\citenamefont {Haldane}(1981)}]{haldane_luttinger_1981}%
  \BibitemOpen
  \bibfield  {author} {\bibinfo {author} {\bibfnamefont {F.~D.~M.}\
  \bibnamefont {Haldane}},\ }\bibfield  {title} {\bibinfo {title} {'{Luttinger}
  liquid theory' of one-dimensional quantum fluids. {I}. {Properties} of the
  {Luttinger} model and their extension to the general 1d interacting spinless
  {Fermi} gas},\ }\href {https://doi.org/10.1088/0022-3719/14/19/010}
  {\bibfield  {journal} {\bibinfo  {journal} {J. Phys. C: Solid State Phys.}\
  }\textbf {\bibinfo {volume} {14}},\ \bibinfo {pages} {2585} (\bibinfo {year}
  {1981})}\BibitemShut {NoStop}%
\bibitem [{\citenamefont {Giamarchi}(2004)}]{giamarchi_quantum_2004}%
  \BibitemOpen
  \bibfield  {author} {\bibinfo {author} {\bibfnamefont {T.}~\bibnamefont
  {Giamarchi}},\ }\href@noop {} {\emph {\bibinfo {title} {Quantum physics in
  one dimension}}}\ (\bibinfo  {publisher} {Clarendon},\ \bibinfo {address}
  {Oxford},\ \bibinfo {year} {2004})\BibitemShut {NoStop}%
\bibitem [{\citenamefont {Matveev}\ and\ \citenamefont
  {Andreev}(2017)}]{matveev_second_2017}%
  \BibitemOpen
  \bibfield  {author} {\bibinfo {author} {\bibfnamefont {K.~A.}\ \bibnamefont
  {Matveev}}\ and\ \bibinfo {author} {\bibfnamefont {A.~V.}\ \bibnamefont
  {Andreev}},\ }\bibfield  {title} {\bibinfo {title} {Second {Sound} in
  {Systems} of {One}-{Dimensional} {Fermions}},\ }\href
  {https://doi.org/10.1103/PhysRevLett.119.266801} {\bibfield  {journal}
  {\bibinfo  {journal} {Phys. Rev. Lett.}\ }\textbf {\bibinfo {volume} {119}},\
  \bibinfo {pages} {266801} (\bibinfo {year} {2017})}\BibitemShut {NoStop}%
\bibitem [{\citenamefont {Matveev}\ and\ \citenamefont
  {Andreev}(2018{\natexlab{a}})}]{matveev_hybrid_2018}%
  \BibitemOpen
  \bibfield  {author} {\bibinfo {author} {\bibfnamefont {K.~A.}\ \bibnamefont
  {Matveev}}\ and\ \bibinfo {author} {\bibfnamefont {A.~V.}\ \bibnamefont
  {Andreev}},\ }\bibfield  {title} {\bibinfo {title} {Hybrid {Sound} {Modes} in
  {One}-{Dimensional} {Quantum} {Liquids}},\ }\href
  {https://doi.org/10.1103/PhysRevLett.121.026803} {\bibfield  {journal}
  {\bibinfo  {journal} {Phys. Rev. Lett.}\ }\textbf {\bibinfo {volume} {121}},\
  \bibinfo {pages} {026803} (\bibinfo {year} {2018}{\natexlab{a}})}\BibitemShut
  {NoStop}%
\bibitem [{\citenamefont {Matveev}\ and\ \citenamefont
  {Andreev}(2018{\natexlab{b}})}]{matveev_propagation_2018}%
  \BibitemOpen
  \bibfield  {author} {\bibinfo {author} {\bibfnamefont {K.~A.}\ \bibnamefont
  {Matveev}}\ and\ \bibinfo {author} {\bibfnamefont {A.~V.}\ \bibnamefont
  {Andreev}},\ }\bibfield  {title} {\bibinfo {title} {Propagation and
  attenuation of sound in one-dimensional quantum liquids},\ }\href
  {https://doi.org/10.1103/PhysRevB.98.155441} {\bibfield  {journal} {\bibinfo
  {journal} {Phys. Rev. B}\ }\textbf {\bibinfo {volume} {98}},\ \bibinfo
  {pages} {155441} (\bibinfo {year} {2018}{\natexlab{b}})}\BibitemShut
  {NoStop}%
\bibitem [{\citenamefont {Landau}(1941)}]{landau_theory_1941}%
  \BibitemOpen
  \bibfield  {author} {\bibinfo {author} {\bibfnamefont {L.~D.}\ \bibnamefont
  {Landau}},\ }\bibfield  {title} {\bibinfo {title} {The {Theory} of
  {Superfluidity} of {Helium} {II}},\ }\href@noop {} {\bibfield  {journal}
  {\bibinfo  {journal} {J. Phys. USSR}\ }\textbf {\bibinfo {volume} {5}},\
  \bibinfo {pages} {71} (\bibinfo {year} {1941})}\BibitemShut {NoStop}%
\bibitem [{\citenamefont {Khalatnikov}(2000)}]{khalatnikov_introduction_2000}%
  \BibitemOpen
  \bibfield  {author} {\bibinfo {author} {\bibfnamefont {I.~M.}\ \bibnamefont
  {Khalatnikov}},\ }\href@noop {} {\emph {\bibinfo {title} {An introduction to
  the theory of superfluidity}}}\ (\bibinfo  {publisher} {Perseus},\ \bibinfo
  {address} {New York},\ \bibinfo {year} {2000})\BibitemShut {NoStop}%
\bibitem [{\citenamefont {Lunde}\ \emph {et~al.}(2007)\citenamefont {Lunde},
  \citenamefont {Flensberg},\ and\ \citenamefont
  {Glazman}}]{lunde_three-particle_2007}%
  \BibitemOpen
  \bibfield  {author} {\bibinfo {author} {\bibfnamefont {A.~M.}\ \bibnamefont
  {Lunde}}, \bibinfo {author} {\bibfnamefont {K.}~\bibnamefont {Flensberg}},\
  and\ \bibinfo {author} {\bibfnamefont {L.~I.}\ \bibnamefont {Glazman}},\
  }\bibfield  {title} {\bibinfo {title} {Three-particle collisions in quantum
  wires: {Corrections} to thermopower and conductance},\ }\href
  {https://doi.org/10.1103/PhysRevB.75.245418} {\bibfield  {journal} {\bibinfo
  {journal} {Phys. Rev. B}\ }\textbf {\bibinfo {volume} {75}},\ \bibinfo
  {pages} {245418} (\bibinfo {year} {2007})}\BibitemShut {NoStop}%
\bibitem [{\citenamefont {Micklitz}\ \emph {et~al.}(2010)\citenamefont
  {Micklitz}, \citenamefont {Rech},\ and\ \citenamefont
  {Matveev}}]{micklitz_transport_2010}%
  \BibitemOpen
  \bibfield  {author} {\bibinfo {author} {\bibfnamefont {T.}~\bibnamefont
  {Micklitz}}, \bibinfo {author} {\bibfnamefont {J.}~\bibnamefont {Rech}},\
  and\ \bibinfo {author} {\bibfnamefont {K.~A.}\ \bibnamefont {Matveev}},\
  }\bibfield  {title} {\bibinfo {title} {Transport properties of partially
  equilibrated quantum wires},\ }\href
  {https://doi.org/10.1103/PhysRevB.81.115313} {\bibfield  {journal} {\bibinfo
  {journal} {Phys. Rev. B}\ }\textbf {\bibinfo {volume} {81}},\ \bibinfo
  {pages} {115313} (\bibinfo {year} {2010})}\BibitemShut {NoStop}%
\bibitem [{\citenamefont {Matveev}\ and\ \citenamefont
  {Andreev}(2012{\natexlab{a}})}]{matveev_equilibration_2012}%
  \BibitemOpen
  \bibfield  {author} {\bibinfo {author} {\bibfnamefont {K.~A.}\ \bibnamefont
  {Matveev}}\ and\ \bibinfo {author} {\bibfnamefont {A.~V.}\ \bibnamefont
  {Andreev}},\ }\bibfield  {title} {\bibinfo {title} {Equilibration of a
  spinless {Luttinger} liquid},\ }\href
  {https://doi.org/10.1103/PhysRevB.85.041102} {\bibfield  {journal} {\bibinfo
  {journal} {Phys. Rev. B}\ }\textbf {\bibinfo {volume} {85}},\ \bibinfo
  {pages} {041102(R)} (\bibinfo {year} {2012}{\natexlab{a}})}\BibitemShut
  {NoStop}%
\bibitem [{\citenamefont {Matveev}\ and\ \citenamefont
  {Andreev}(2012{\natexlab{b}})}]{matveev_scattering_2012}%
  \BibitemOpen
  \bibfield  {author} {\bibinfo {author} {\bibfnamefont {K.~A.}\ \bibnamefont
  {Matveev}}\ and\ \bibinfo {author} {\bibfnamefont {A.~V.}\ \bibnamefont
  {Andreev}},\ }\bibfield  {title} {\bibinfo {title} {Scattering of hole
  excitations in a one-dimensional spinless quantum liquid},\ }\href
  {https://doi.org/10.1103/PhysRevB.86.045136} {\bibfield  {journal} {\bibinfo
  {journal} {Phys. Rev. B}\ }\textbf {\bibinfo {volume} {86}},\ \bibinfo
  {pages} {045136} (\bibinfo {year} {2012}{\natexlab{b}})}\BibitemShut
  {NoStop}%
\bibitem [{\citenamefont {Abrikosov}(1988)}]{abrikosov_fundamentals_1988}%
  \BibitemOpen
  \bibfield  {author} {\bibinfo {author} {\bibfnamefont {A.~A.}\ \bibnamefont
  {Abrikosov}},\ }\href@noop {} {\emph {\bibinfo {title} {Fundamentals of the
  {Theory} of {Metals}}}}\ (\bibinfo  {publisher} {North-Holland},\ \bibinfo
  {address} {Amsterdam},\ \bibinfo {year} {1988})\BibitemShut {NoStop}%
\bibitem [{one()}]{onefootnote}%
  \BibitemOpen
  \href@noop {} {}\bibinfo {note} {Henceforth we assume normal ordering of
  bosonic fields.}\BibitemShut {Stop}%
\bibitem [{\citenamefont {DeGottardi}\ and\ \citenamefont
  {Matveev}(2015)}]{degottardi_electrical_2015}%
  \BibitemOpen
  \bibfield  {author} {\bibinfo {author} {\bibfnamefont {W.}~\bibnamefont
  {DeGottardi}}\ and\ \bibinfo {author} {\bibfnamefont {K.~A.}\ \bibnamefont
  {Matveev}},\ }\bibfield  {title} {\bibinfo {title} {Electrical and {Thermal}
  {Transport} in {Inhomogeneous} {Luttinger} {Liquids}},\ }\href
  {https://doi.org/10.1103/PhysRevLett.114.236405} {\bibfield  {journal}
  {\bibinfo  {journal} {Phys. Rev. Lett.}\ }\textbf {\bibinfo {volume} {114}},\
  \bibinfo {pages} {236405} (\bibinfo {year} {2015})}\BibitemShut {NoStop}%
\end{thebibliography}
\end{document}